\documentclass[twoside]{article}
\usepackage{fleqn,espcrc2}
\usepackage{graphicx}
\usepackage{amsmath}
\usepackage{amsfonts}
\usepackage{amssymb}
\usepackage{amsthm}

\title{The powers of deconfinement 
\thanks{Presented by E.~Meg\'{\i}as at
    the 14th International Conference In Quantum Chromodynamics (QCD
    08) 7-12th Jul 2008, Montpellier, France. Work supported by the
    Fulbright Program of the U.S. Department of State and Spanish
    MEC, the Spanish DGI and FEDER funds
    with grant FIS2005-00810, Junta de Andaluc\'{\i}a grant
    FQM-225-05, EU Integrated Infrastructure Initiative Hadron Physics
    Project contract RII3-CT-2004-506078, and U.S. Department of
    Energy contract DE-AC02-98CH10886.  }}

\author{E.~Meg\'{\i}as\address{Physics Department, Brookhaven National
    Laboratory, Upton, New York 11973, USA}, 
E.~Ruiz Arriola\address[UGR]{Departamento de F\'{\i}sica At\'omica,
  Molecular y Nuclear, Universidad de Granada, E-18071 Granada, Spain}
and L.L.~Salcedo\addressmark[UGR]} 
       
\begin{document}

\begin{abstract}
The trace anomaly of gluodynamics encodes the breakdown of classical
scale invariance due to interactions around the deconfinement phase
transition. While it is expected that at high temperatures
perturbation theory becomes applicable we show that current lattice
calculations are far from the perturbative regime and are dominated
instead by inverse even power corrections in the temperature, while
the total perturbative contribution is estimated to be extremely small
and compatible with zero within error bars.  We provide an
interpretation in terms of dimension-two gluon condensate of the
dimensionally reduced theory  which value agrees with a similar
analysis of power corrections from available lattice data for the
renormalized Polyakov loop and the heavy quark-antiquark free energy
in the deconfined phase of QCD~\cite{Megias:2005ve,Megias:2007pq}.
\end{abstract}

\maketitle


{\bf Introduction.} The Lagrangian of gluodynamics is conformal
invariant, reflecting the absence of an explicit scale.  The
divergence of the dilatation current equals the trace of the improved
energy-momentum tensor $\Theta^\mu_\mu$~\cite{Callan:1970ze} and
vanishes classically. Quantum-mechanically yields instead the
so-called {\it ``trace anomaly''}~\cite{Collins:1976yq}. It reflects
the breaking of scale invariance which introduces a single mass scale,
$\Lambda_{\rm QCD}$. The dimensionless {\it ``interaction measure''}
$\Delta =T\partial_T (p/T^4)=(\epsilon - 3 p )/T^4$
quantifies the departure from the conformal limit $\epsilon = 3p$,
which corresponds to a gas of free massless particles. At finite
temperature, the energy density $\epsilon$ and the pressure $p$ enter
as~\cite{Landsman:1986uw,Ellis:1998kj,Drummond:1999si,Agasian:2001bj},
\begin{equation}
T^4 \, \Delta   \equiv \epsilon - 3p 
= \frac{\beta(g)}{2g} 
\langle (G^a_{\mu\nu})^2\rangle \equiv \langle \Theta^\mu_\mu\rangle  \,,
\label{eq:tr_an}
\end{equation}
where $G_{\mu\nu} = \partial_\mu A_\nu - \partial_\nu A_\mu - ig
[A_\mu,A_\nu]$ is the field strength tensor and $\beta(g) = \mu
\partial g / \partial \mu = - b_0 g^3 + {\cal O}(g^5) $ is the
beta function, with $b_0 = 11 N_c/(48 \pi^2) $. A good knowledge
of $\Delta$ is crucial to understand the deconfinement process, where
the non perturbative (NP) nature of low energy QCD seems to play a
prominent role \cite{Lichtenegger:2008mh}. In this contribution we
analyze the highly NP behaviour of the trace anomaly just above the
phase transition and describe it in a way that is consistent with
other thermal observables (see~\cite{Megias:2008ip} for further
details).


{\bf Low and high temperatures.} At low temperatures $\Delta$ is
dominated by the lightest confined states in the spectrum. In
gluodynamics the lightest glueball mass
$m_G \approx 1.3\, {\rm GeV}$ is much heavier than $T_c \approx 270\,
{\rm MeV}$, and the pressure is $p \sim e^{-m_G /T}$, so
$\Delta \sim  e^{-m_G/T}  \, , \quad  T \ll T_c \, ,
$
indicating an exponentially small violation of scale
invariance.\footnote{In full QCD for massless quarks one has a gas of
  weakly interacting massless pions and $\Delta \sim T^4 /
  \Lambda_{\rm QCD}^4 $ as dictated by chiral
  symmetry~\cite{Gerber:1988tt}.} At very high temperatures one also
expects scale invariance to be restored while asymptotic freedom
guarantees the applicability of perturbative QCD (pQCD). Actually,
from the pressure to two loop one has~\cite{Kapusta:1979fh}
\begin{eqnarray}
\Delta= \frac{N_c(N_c^2 -1)}{72 }
b_0 g^4(\mu) + {\cal O} ( g^5 ) \, , \quad  T \gg T_c  \, 
\label{eq:pt}
\end{eqnarray} 
where $1/g^2(\mu)= b_0 \log( \mu^2/ \Lambda_{\rm QCD}^2)$. It should
be noted the ambiguity in this result, since generally one has both
the temperature $T$ and the ($\overline{\rm MS}$)-renormalization
scale, $\mu$, for which one takes the reasonable but {\it arbitrary}
choice $\mu \sim 2 \pi T - 4 \pi T$.  Higher order corrections
including up to $ g^6 \log g $ can be traced
from~\cite{Kajantie:2002wa}. The infrared problems of the perturbative
expansion yield poor convergence at the lattice QCD available
temperatures $T < 6 \,T_c $. Perturbation theory contains only
logarithms in the temperature, suggesting a mild temperature
dependence.  This feature is shared by hard thermal loops (HTL) and
other resummation techniques of infrared divergencies (see
e.g.~\cite{Andersen:1999sf,Andersen:2004fp}).  Actually, the value
they find $\Delta_{\rm HTL} =0 \pm 0.5 $ for $T > T_c$ is compatible
with zero within uncertainties. Furthermore, it is not clear at {\it
  what} temperatures is the pQCD result dominating $\Delta$.

\medskip

\begin{figure*}[ttt]
\begin{center}
\includegraphics[width=7.3cm,height=5cm]{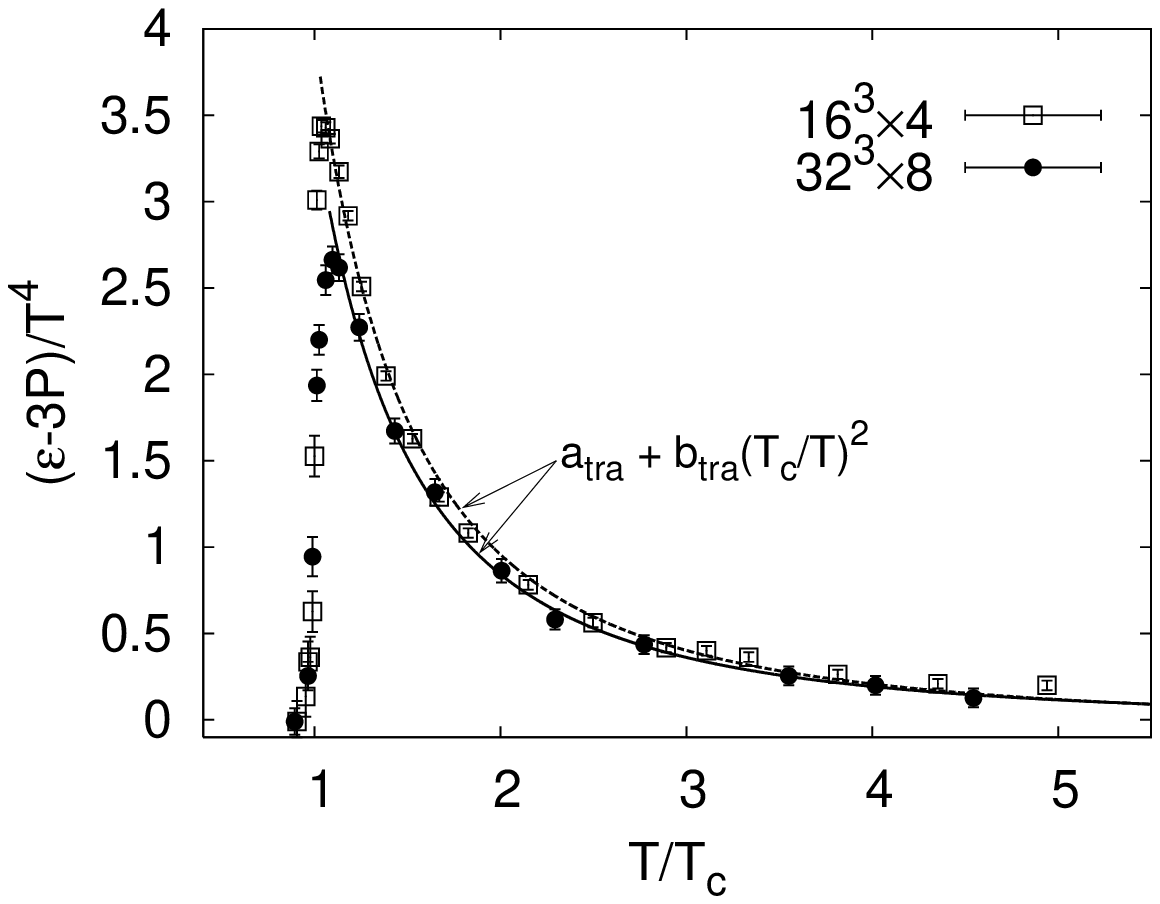}
\includegraphics[width=7.3cm,height=5cm]{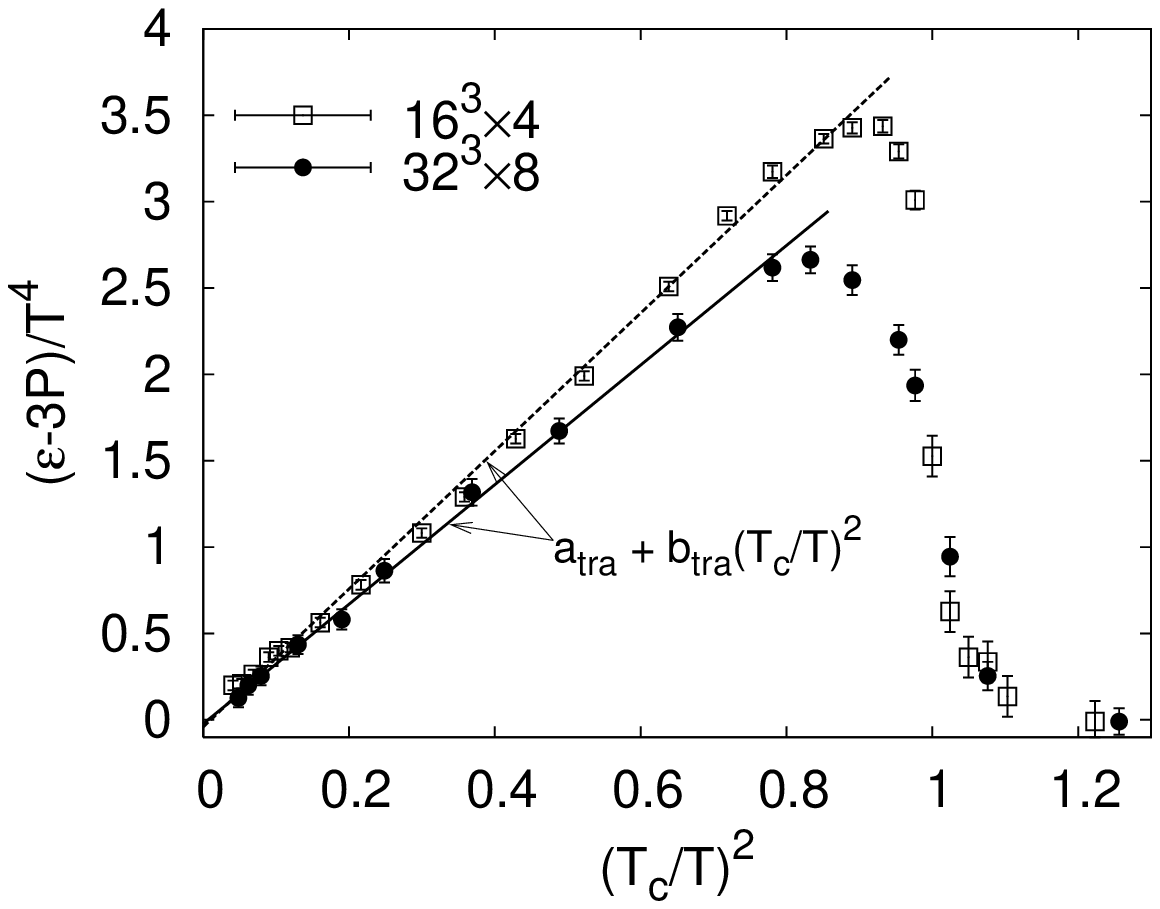}
\end{center}
\caption{\small Trace anomaly $ (\epsilon -3 p)/T^4 $ as a function of
$T$ (left) and $1/T^2$ (right) (in units of $T_c$). Lattice data are
from~\cite{Boyd:1996bx} for $N_\sigma^3 \times N_\tau = 16^3 \times 4$
and $32^3 \times 8$. The fits use Eq.~(\ref{eq:e3pfit}) with $a_{\rm
tra}$ and $b_{\rm tra}$ adjustable constants.}
\label{fig:e3p}
\end{figure*}

{\bf Thermal power corrections in gluodynamics.}  The interaction
measure on the lattice for gluodynamics~\cite{Boyd:1996bx} is shown in
Fig.~\ref{fig:e3p} and, as expected, is very small below $T_c$.  It
increases suddenly near and above $T_c$ by latent heat of
deconfinement, and raises a maximum at $T \approx 1.1\, T_c$. Then it
has a gradual decrease reaching small values at $T=5\, T_c$. The high
value of $\Delta$ for $ T_c < T < (2.5-3) T_c$ corresponds to a
strongly interacting Quark-Gluon Plasma picture. 

In previous works~\cite{Megias:2005ve,Megias:2007pq} we have detected
the presence of inverse power corrections in other thermal observables.  In
Fig.~\ref{fig:e3p} we plot $(\epsilon - 3p)/T^4$ as a function of
$1/T^2$ (in units of $T_c$) exhibiting an obvious straight line
behaviour in the region slightly above the critical temperature, 
\begin{equation}
\Delta_{\rm latt} =
a_{\rm tra} + b_{\rm tra} \left( T_c /T \right)^2 \, \, , 
\label{eq:e3pfit} 
\end{equation}  
and corresponding to a ``power correction'' in temperature.  A fit of
the lattice data ($N_\sigma^3 \times N_\tau = 32^3 \times 8$) for $
1.13 T_c \le T \le 4.54 T_c $ yields $a_{\rm tra} = -0.02(4)\,,\;
b_{\rm tra} = 3.46(13)\,, \; \chi^2/ {\rm DOF} = 0.35 \,$. Power
corrections also appear in $\epsilon$ and $p$, just by applying the
standard thermodynamic relations. Note that if the power correction
was absent the transition towards the free gas value would take over
much faster indicating a weakly interacting quark-gluon plasma in the
neighbourhood of $T_c$.

Pisarski~\cite{Pisarski:2006yk} has suggested to interpolate between
$T=T_c$ where $p=0$ for non-interacting and heavy glueballs and the
free massless gluon gas value $p \sim T^4$ at $T \gg T_c$. If we choose
\begin{eqnarray}
p= \frac{(N_c^2-1) \pi^2}{45} T^4 \left[ 1- \left( \frac{T_c}{T}
  \right)^n \right] \, , \qquad T \ge T_c \, , 
\end{eqnarray} 
with $n$ arbitrary we get 
\begin{eqnarray} 
\Delta = \frac{n(N_c^2-1) \pi^2}{45} \left( \frac{T_c}{T} \right)^n \,,
\qquad T \ge T_c \, , 
\label{eq:e3pnfit}
\end{eqnarray} 
which for $n=2$ corresponds to take $a=0$ and $b=3.51$ in
Eq.~(\ref{eq:e3pfit}), in excellent agreement with the fit to the
lattice data. This thermodynamic consistency does not explain however
{\it why} there is a power correction with $n=2$. A fit of
Eq.~(\ref{eq:e3pnfit}) to the data of Fig.~\ref{fig:e3p} for the same
range as in Eq.~(\ref{eq:e3pfit}), yields $n=1.97(5)$ with
$\chi^2/{\rm DOF}=0.62$, indicating the robustness of the power.

The lattice behaviour of $\Delta$ clearly contradicts perturbation
theory~\cite{Kapusta:1979fh,Kajantie:2002wa} and resummations
thereof~\cite{Andersen:1999sf,Andersen:2004fp} explaining why they
have flagrantly failed to describe the data of the free energy below
$3\, T_c$. Our discussion above shows that these approaches would
yield a powerless contribution, $\Delta_{\rm PT}$, which should
ultimately be identified with the {\it almost} constant and vanishing
$a_{\rm tra}$ of Eq.~(\ref{eq:e3pfit}) rather than with the full
result from the lattice~\cite{Boyd:1996bx}.  Actually, the maximum
lattice temperature $T=5\, T_c$ may still be far from the pQCD
estimate since the power correction provides the bulk of the full
result at this temperature. For $N_c=3$ the ${\cal O} (g^5) $
correction to Eq.~(\ref{eq:pt}) corresponds to multiply it by $(1-6
g(\mu)/\pi)$~\cite{Kajantie:2002wa} which becomes small for $ g(\mu) \ll
\pi /6 $ or $\mu \gg 10^{11} \Lambda_{\rm QCD}$. This delayed onset of
pQCD is not new and happens in the study of exclusive processes at
high energies (see e.g. Ref.~\cite{RuizArriola:2008sq}).

\medskip

{\bf Dimension two gluon condensate.} Power corrections are
unmistakable high energy traces of NP low energy effects. Within QCD
sum rules power corrections at high $Q^2$ are usually related to {\it
  local} condensates as suggested by the OPE. 
The gluon condensate $\langle G^2
\rangle \equiv g^2 \langle (G^a_{\mu\nu})^2\rangle $ describes the
anomalous (and not spontaneous) breaking of scale invariance, and
hence is not an order parameter of the phase transition. Actually, the
order parameter is the vacuum expectation value of the Polyakov loop $L(T)$
which signals the breaking of the ${\mathbb Z}(N_c)$ discrete symmetry
of gluodynamics as well as the deconfinement transition. A dimension
two gluon condensate naturally appears from a computation of $L(T)$
which in the static gauge, $\partial_0 A_0 ({\mathbf x},x_0)=0$,
in a  Gaussian-like, large $N_c$ motivated, approximation
gives~\cite{Megias:2005ve}
\begin{equation}
\left\langle \frac{1}{N_c} \, {\rm tr}_c \, e^{i g A_0({\mathbf x})/T } \right\rangle = 
\exp \left[  -\frac{g^2 \langle A_{0,a}^2 \rangle }{4N_c T^2} \right] 
\,,
\label{eq:defPL}
\end{equation}
valid up to ${\cal O}(g^5)$ in pQCD. $A_0$ is the gluon field in the
(Euclidean) time direction.

The dimension two gluon condensate $g^2\langle A_{0,a}^2 \rangle$ is
obtained from the gluon propagator of the dimensionally reduced
theory, $D_{00}$, by taking the coincidence limit. The perturbative
propagator $D_{00}^{\rm P}({\mathbf k}) = 1/({\mathbf k}^2 + m_D^2) +
{\cal O}(g^2)$, being $m_D \sim T$ the Debye mass, leads to the known
perturbative result~\cite{Gava:1981qd} and fails
to reproduce lattice data below $6\, T_c$~\cite{Kaczmarek:2002mc}. A
NP model is proposed in Ref.~\cite{Megias:2005ve} to describe the
lattice data of the Polyakov loop, and it consists in a new piece in
the gluon propagator driven by a positive mass dimension parameter:
\begin{equation}
D_{00}({\mathbf k}) = D_{00}^{\rm P}({\mathbf k}) + D_{00}^{\rm NP}({\mathbf k}) \,, 
\label{eq:NPmodel}
\end{equation}
where $D_{00}^{\rm NP}({\mathbf k}) =m_G^2 /({\mathbf k}^2 + m_D^2)^2 $.
This ansatz parallels a zero temperature one~\cite{Chetyrkin:1998yr},
where the dimension two condensate provides the short-distance NP
physics of QCD and  at zero temperature this contribution yields the well
known NP linear term in the $\overline{q}q$ potential. 
A justification of Eq.~(\ref{eq:NPmodel}) based on Schwinger-Dyson
methods has been given~\cite{Gogokhia:2005gk}. 
The Gaussian approximation has also been used in
Ref.~\cite{Megias:2007pq} to compute the singlet free energy of a
heavy ${\overline q}q$ pair~\cite{Kaczmarek:2002mc,Kaczmarek:2004gv},
through the correlation function of Polyakov loops.

\medskip

{ \bf Non perturbative contribution to the Trace Anomaly.} The model
of Eq.~(\ref{eq:NPmodel}) can easily be used to compute the trace
anomaly Eq.~(\ref{eq:tr_an}) in gluodynamics. Assuming the leading NP
contribution to be encoded in the $A_{0,a}$ field and taking
$A_{i,a}=0$ yields~\cite{Megias:2008dv}
\begin{equation}
\langle (G^a_{\mu\nu})^2\rangle^{\rm NP} 
= -6 m_D^2 \langle A_{0,a}^2\rangle^{\rm NP}\,. 
\end{equation}  
Note that the NP model is formulated in the dimensionally reduced
theory, so the gluon fields are static. This formula produces the
thermal power behaviour of Eq.~(\ref{eq:e3pfit}) with
\begin{equation}
b_{\rm tra} T_c^2 = 
-(3\beta(g)/ g) \hat{m}_D^2  \langle A_{0,a}^2\rangle^{\rm NP}  \,,
\label{eq:btra}
\end{equation}
where ${\hat m}_D \equiv m_D/T$.  If we consider the perturbative
value of the beta function $\beta(g) \sim g^3 + {\cal O}(g^5) $, the
r.h.s. of Eq.~(\ref{eq:btra}) shows a factor $g^2$ in addition to the
dimension two gluon condensate $g^2 \langle A_{0,a}^2\rangle^{\rm
  NP}$. So the fit of the trace anomaly data is sensitive to the value
of the smooth $T$-dependent $g$, without jeopardizing the power
correction. 
When we consider the perturbative
value $g_P$ up to 2-loops, we get from the fit of the trace anomaly
a value of $g^2 \langle A_{0,a}^2\rangle^{\rm NP}$
which is a factor $1.5$ smaller than from other observables. This
disagreement could be partly explained on the basis of certain
ambiguity of $g$ in the NP regime. A better fit of the Polyakov loop
and heavy quark free energy lattice data in the regime $T_c < T < 4\,
T_c$ is obtained for a slightly smaller $g$ than $g_P$, i.e. $g = 1.26
- 1.46$~\cite{Megias:2007pq}. Taking this value we get from
Eq.~(\ref{eq:btra}) $g^2 \langle A_{0,a}^2 \rangle^{\rm NP} = (2.86
\pm 0.24 \, T_c)^2$, a better overall agreement, see
Table~\ref{tab:1}. From Renormalization Group requirements it is
possible to observe that the consistency of the model is warrantied if
the coupling constant $\alpha_s(\mu) \equiv g^2(\mu)/(4\pi) $ has a
behaviour $\sim 1/\mu^2$ at low enough
temperature~\cite{Megias:2008ip}.  This behaviour is just what is
obtained within the Analitic Perturbation Theory formalism, after
extracting the Landau pole~\cite{Shirkov:2008hf}, with the
corresponding decrease of the perturbative value of $\alpha_s(\mu)$.
This decrease could approximately explain the best-fit value of~$g$.

While there might exist other explanations to the observed thermal
power corrections our results of Table~\ref{tab:1} suggest 
an unified and coherent description of observables in the non
perturbative regime of the deconfined phase (sQGP) in terms of the
dimension two gluon condensate~\cite{Megias:2008ip}.


\begin{table}[htb]
\begin{center}
\begin{tabular}{|c|c|}
\hline
{\bf Observable} &  $\hspace{0.05cm} {\bf g^2 \langle A_{0,a}^2 \rangle^{\rm NP} }$ \hspace{0.05cm}  \\
\hline
\hspace{0.05cm} Polyakov loop~\cite{Megias:2005ve} \hspace{0.05cm} &  \hspace{0.05cm} $(3.22 \pm 0.07 \, T_c)^2   $ \hspace{0.05cm} \\
\hspace{0.05cm} Heavy $\overline{q}q$ free energy~\cite{Megias:2007pq} \hspace{0.05cm}  &  \hspace{0.05cm} $(3.33 \pm 0.19 \, T_c)^2$  \hspace{0.05cm} \\ 
\hspace{0.05cm} Trace Anomaly  \hspace{0.05cm} &   \hspace{0.05cm} $(2.86 \pm 0.24 \, T_c)^2$ \hspace{0.05cm} \\ 
\hline
\end{tabular}
\end{center}
\caption{\small Values (in units of $T_c$) of the dimension two gluon
  condensate from a fits in the deconfined phase of gluodynamics:
  Polyakov loop, singlet free energy of heavy quark-antiquark and
  trace anomaly. Lattice data are with $N_\tau = 8$. The error
  reflects an uncertainty in the coupling constant $g=1.26-1.46$,
  being the highest value the perturbative $g_P$ up to 2-loops at $T =
  2 \, T_c$. The critical temperature in gluodynamics is $T_c = 270
  \pm 2 \,$MeV~\cite{Kaczmarek:2002mc}.}
\label{tab:1}
\end{table}

\medskip




%




\end{document}